\documentclass[a4papers,12pt]{article}
\usepackage{graphicx}% Include figure files
\usepackage{dcolumn} % Align table columns on decimal point
\usepackage{bm}      % bold math
\usepackage{amsmath}
\usepackage{amsfonts}
\usepackage{amssymb}
\newcommand{\vett}[1]{\mathbf{#1}}
\newcommand{\vir}{\mathcal{V}}

\newcommand{\dqj}{ {\dot q}^{(j)} }

\newcommand{\eff}{\mathrm{eff}}
\newcommand{\ret}{\mathrm{ret}}

\newcommand{\Name}[1]{\textsc{#1},\ }
\newcommand{\REVIEW}[4]{\textit{#1}, \textbf{#2}, #4 (#3)}
\newcommand{\Book}[1]{\textit{#1}, }
\newcommand{\Publ}[1]{(#1)}
\newcommand{\Year}[1]{#1}
\newcommand{\Editor}[1]{#1 eds.,}
\newcommand{\Page}[1]{, #1}
\newcommand{\Vol}[1]{#1, }
\title{Discrete Matter,  Far Fields, and  Dark Matter}
\author{A. Carati\footnote{Department of Mathematics, University  of  Milano -
                Via Saldini 50, I-20133 Milano, Italy. } \and S.L.
		Cacciatori\footnote{Department of Physical and Mathematical Sciences,
		Insubria University - Via Valleggio 11, I-22100 Como, Italy.} \and
		L. Galgani$^*$ } 
\date{\today}% It is always \today, today,
\begin{document}

\maketitle

\abstract{
We show that in cosmology the gravitational action of the far away
matter has quite relevant effects, if retardation of the forces and
discreteness of matter (with its spatial correlation) are taken into
account.  The expansion rate is found to be determined by the density
of the far away matter, i.e., by the density of matter at remote
times. This leads to the introduction of an effective density, which
has to be five times larger than the present one, if the present
expansion rate is to be accounted for.  The force per unit mass on a
test particle is found to be of the order of $0.2\, cH_0$. The
corresponding contribution to the virial of the forces for a cluster
of galaxies is also discussed, and it is shown that it fits the
observations if a decorrelation property of the forces at two
separated points is assumed. So it appears that the gravitational
effects of the far away matter may have the same order of magnitude as
the corresponding local effects of dark matter. }

\section{Introduction}
In most cosmological models usually considered, the matter is dealt
with as a continuous medium. The aim of the present paper is to point
out how relevant are the gravitational effects which are due to the
discreteness of matter, thought of as constituted of galaxies
described as point particles, if one takes into account both the role
of retardation of the forces (as required by general relativity) and
the correlated nature of the positions of the galaxies.  Concerning
the role of discreteness, the key point is that the gravitational
force on a test particle due to a continuous matter with a spherically
symmetric density vanishes. Instead, for a matter constituted of point
particles whose positions are dealt with as random variables with a
spherically symmetric probability distribution, it is only the mean
gravitational force that vanishes, while the fluctuations can be very
large. This actually is the point of view that was taken by
Chandrasekhar and von Neumann in connection with the motions of stars
(see the review \cite{chandra}).  They showed that, if the positions
of the stars about a test particle are considered as (independent)
random variables, then the force on the test particle may be very
large; actually, this happens with so huge a probability that the
variance of the force is even divergent. It will be shown here that
very large fluctuations of the force on a test particle occur also in
the case of galaxies. However, while in the case of stars this is due
to the occurrence of close encounters, in the case of galaxies the
largeness of the fluctuations is instead due to the gravitational
contribution of the far galaxies, when one takes into account both the
retarded character of their action and the correlated nature of the
positions of the galaxies.

A probabilistic approach in a cosmological context, with galaxies
described as point particles, whose positions are dealt with as random
variables presenting correlations, is a familiar one since a rather
long time; see for example the book of Mandelbrot \cite{mandel}, the
book \cite{peebles} by Peebles and the work \cite{dp} by Davis and
Peebles.  Particular emphasis on the possible fractal nature of matter
distribution was given, in addition to Mandelbrot, by several
authors. See for example the reviews \cite{sylos} by Sylos Labini et
al., \cite{coleman} by Coleman and Pietronero and \cite{combes} by
Combes, and the works \cite{remo} by Ruffini et al., \cite{roma2} and
\cite{roma3} by Gabrielli at al. and finally the work \cite{roma} by
Joyce et al.  Now, in all such papers the nonrelativistic
approximation for gravitation was used, and retardation was altogether
neglected, so that one is actually dealing with purely static
Newtonian gravitational forces.

The main original contribution of the present paper consists in
showing that, if retardation is taken into account (together with
Hubble's law and the correlated nature of the positions of the
galaxies), then the gravitational action of far away matter enters the
game and may, in some cases, be the dominant one.

This will be shown by considering an extremely simplified model of the
Universe, with the Hubble constant hold fixed at its present value
$H_0$. Two results will be obtained.  First we show that the influence
of the far away galaxies can be described as corresponding to the
existence of an effective density which is about five times larger
than the present barionic one, i.e., about equal to the usually
estimated density of dark matter. Then we look at the force on a test
particle. We show that, if the correlated nature of the positions of
the galaxies is taken into account, the force (per unit mass) can be
estimated as given by $0.2\, c H_0$ ($c$ being the speed of light),
which is about the value of the acceleration at which the influence of
dark matter starts to be felt.  Such results thus appear to indicate
that far away matter may produce gravitational effects comparable to
those usually attributed to local dark matter.

We finally give a preliminary discussion of the problem whether such
an estimate of the gravitational action of far away matter may
account, through the virial theorem, for the observed velocity
dispersion in clusters of galaxies. We show that this is possible,
provided the gravitational force of far away matter has a suitable
property concerning its dependence on position. Namely, the force
should not be smooth, and its values at two separated points should
rather be uncorrelated. We point out how the extremely simplified
model here considered may not suffice to settle the question whether
such a decorrelation property should hold, because the answer may
require the introduction of a more realistic model, in which the time
variation of Hubble's constant is taken into account. We leave the
discussion of this interesting point for future work, and in the
present paper we limit ourselves to exhibit, through the simplest
conceivable model, how relevant the role of far away matter may be for
cosmology, if retardation of the forces (in addition to the correlated
nature of the positions of the galaxies) is taken into account.

\section{Definition of the model}
In order to fully take the discrete character of matter into account,
one should in principle deal with an $N$--body problem, in which each
particle is coupled to the gravitational field through the Einstein
equation having all the other particles as sources. This is however a
formidable task.  So we introduce first of all the approximation in
which one looks at the motion of a test particle, when the motion of
the sources is assigned, as given by observational cosmology. This
will naturally lead to a compatibility problem, because the test
particle too will have to move according to the same law. It will be
shown how this compatibility condition is solved through the
introduction of a suitable effective density.

As the simplest model for the motion of the sources, we take a
velocity field which corresponds to Hubble's law, i.e., we neglect
altogether the peculiar velocities (a further comment on this point
will be given later).  Taking a locally Minkowskian coordinate system
centered about an arbitrary point, a particle with position vector
$\vett q$ will then have a velocity
\begin{equation}\label{hubble}
\dot{\vett q}=H_0 \, \vett q\ .
\end{equation}
For the sake of simplicity of the model, the Hubble constant $H_0$
will be assumed to be independent of time. On this point we will come
back later on.  It is easily established that the chart has a local
Hubble horizon $R_0=c/H_0$, where the galaxies have the speed of
light.  Notice furthermore that the form (\ref{hubble}) of Hubble's
law is the one appropriate to our choice of Minkowskian coordinates.
For example, one could choose, as Davis and Peebles (but not Joyce et
al.) do, coordinates \emph{``expanding with the background
cosmological model''}, with respect to which the galaxies have zero
velocity (the peculiar velocities having been neglected); see formula
(1), page 426, of \cite{dp}.  Our choice of coordinates is perhaps
more convenient in the present case, but obviously, just by
definition, the results do not depend on the choice of the coordinates
at all.

So we investigate the gravitational action due to a system of $N$
galaxies whose motions $\vett q_j(t)$, $j=1,\ldots ,N$, are assigned.
The energy--momentum tensor $T^{\mu\nu}$ then is
\begin{equation}\label{timunu}
T^{\mu\nu}=\sum_{j=1}^N \frac 1 {\sqrt g}\ \frac {M_j}{\gamma_j}\
\delta (\vett x-\vett q_j) \dot{\vett q}^\mu_j \dot{\vett q}^\nu_j
\end{equation}
where $M_j$, and $\gamma_j$ are the mass and the Lorentz factor of the
$j$--th particle, $g$ is the determinant of the metric tensor (which
is considered as an unknown of the problem), $\delta$ the Dirac delta
function, and the dot denotes derivative with respect to proper time
along the worldline of the source.  The velocities of the galaxies are
assumed to satisfy Hubble's law (\ref{hubble}), while their position
vectors $\vett q_j$ are considered as random variables, whose
statistical properties will be discussed later.

\section{The perturbation approach}
The study of the solutions of Einstein's equations with the
energy--momentum tensor (\ref{timunu}) as a source still is a
formidable task, and so we limit ourselves to a perturbation approach,
in which the energy--momentum tensor $T^{\mu\nu}$ (\ref{timunu}) is
considered as a perturbation of the vacuum. Following the standard
procedure (see \cite{einstein} or \cite{wein}), we have to determine a
zero--th order solution (the vacuum solution), and solve the Einstein
equations, linearized about it.  The simplest consistent zero--th
order solution is the flat metric, because it will be shown that,
coherently, the perturbation turns out to be small (at least if the
free parameters are chosen in accordance with the observations). We
did not investigate whether there do exist other \emph{ansatzs} for
the vacuum which give better results.  Some further comments on the
perturbation procedure will be given later.

Thus the metric tensor $g_{\mu \nu}$ is written as a perturbation of
the Minkowskian background $\eta_{\mu \nu}$, namely, as $g_{\mu
\nu}=\eta_{\mu \nu} +h_{\mu \nu}$, and it is well known that in the
linear approximation the perturbation $h_{\mu\nu}$ has to satisfy
essentially the wave equation with $T^{\mu\nu}$ as a source. More
precisely, one gets
\begin{equation}\label{aaaa}
\square \big[ h_{\mu\nu}-\frac 12 \eta_{\mu\nu}h\big]= -\frac{16\pi
  G}{c^4} T_{\mu\nu}\ ,
\end{equation}
where $G$ is the gravitational constant, $h$ the trace of
$h_{\mu\nu}$, and $\square=(1/c^2)\partial^2_t-\Delta_2$.  The
solutions are the well known retarded potentials
\begin{equation}
\label{campo}
h_{\mu\nu}=\frac {-2 G}{c^4 }\, \, \sum_{j=1}^N \frac {
M_j}{\gamma_j}\, \left. \frac {2\dqj_\mu \dqj_\nu -c^2\eta_{\mu\nu}}
{|\vett x-{\vett q}_j|}\right|_{t=t_{\ret}} \
\end{equation}
(with ${\vett q}^{(j)}\equiv {\vett q}_j$).

\section{The mean metric, the compatibility condition and the 
effective density}
In order to implement in a suitable sense the compatibility condition
previously mentioned, 
we  now  make reference to the mean metric,
which is obtained by averaging over the position vectors of the
galaxies, considered   as random variables. 
For  a spherically symmetric probability distribution it is immediately seen
that the mean of each of the off--diagonal terms vanishes, and that
the means of the spatial diagonal components are all equal.
Denoting the mean by
$\langle\, .\, \rangle$, the  mean metric at the origin is then 
$$
ds^2= \langle\ {g}_{\mu\nu}\ \rangle\  dx^\mu dx^\nu=
(1-\alpha-3\beta)\, 
c^2 dt^2 -
(1+\alpha+\beta) dl^2
$$
where $dl^2=dx^2+dy^2+dz^2$ and
\begin{equation}\label{coef}
\alpha=\frac {2G}{c^2}\ \langle\ \sum_j \frac {M_j}{|\vett q_j|}\
\rangle \ ,\quad
\beta \, \raisebox{-0.5ex}{$\stackrel{<}{\sim}$} \frac {4GH_0^2}{3c^4}\  
\langle\ \sum_j M_j {|\vett q_j|}\ \rangle\ . 
\end{equation} 
This actually is a spatially flat Friedmann--Robertson--Walker
metric. We can now formulate the compatibility
condition as the requirement that  the expansion rate corresponding
to such a metric  coincide with the one ($H_0$) that was assumed for the
sources. The condition then takes the form
\begin{equation}\label{consistenza}
\frac 12\frac {d}{dt} \log \frac {1+\alpha+\beta}{1-\alpha-3\beta}=H_0\ .
\end{equation}

The sums (\ref{coef}) defining the coefficients $\alpha$ and $\beta$
might be estimated through integrals involving a suitable effective
matter density. There arises however the problem that, due to the
retarded character of the time entering the expressions for $\alpha$
and $\beta$, the galaxies lying near the border of the chart are to be
taken at times near that of the big bang, at which the density
diverges.  This by the way shows that only the galaxies at the border
are the relevant ones. This very fact, however, also allows one to
solve the problem just mentioned, because one can then introduce an
effective density $\rho_{\eff}$ having the property that both
relations
$$
\langle\ \sum \frac {M_j}{|\vett q_j|}\ \rangle
\simeq 4\pi
\rho_{\eff} \ \frac{{R_0}^2}2 \ , \quad
\langle\ \sum M_j {|\vett q_j|}\ \rangle\  \simeq 4\pi \rho_{\eff}
\frac{{R_0}^4}4 \ 
$$ 
hold, with the same effective density.
This gives
\begin{equation}\label{aaa}
\alpha\simeq {4\pi G} \rho_{\eff} {{R_0}^2}/{c^2}\ ,\quad
\beta\ <  (2/3) \alpha\ .
\end{equation}
Using ${\dot R}_0=c$, one then gets
$$
\dot \alpha\simeq \frac {8\pi G}{c^2} \ \rho_{\eff}\  R_0c\ , \quad
\dot \beta \simeq \frac 23 \dot\alpha \ .
$$
With  these expressions for $\dot \alpha$ and $\dot \beta$,  the consistency 
condition (\ref{consistenza}) then becomes an algebraic one, which gives
for $\rho_{\eff}$ the  value  
\begin{equation}\label{rhoeff}
\rho_{\eff}\simeq \frac 14\,  \frac {3H_0^2}{8\pi G}\simeq  5 \rho_0 \ ,
\end{equation}
where $\rho_0=\Omega_0\, \big({3H_0^2})/\big({8\pi G})$, with 
$\Omega_0\simeq 0.05$, is the  observed barionic density at the present
time. Notice by the way that the perturbation procedure appears to be
qualitatively consistent, because the first--order perturbation turns
out to be small, of the order of one tenth  the unperturbed one. 

This is the first result of the present paper. Due to the retarded
nature of the potentials, it turns out that the far away galaxies are
the ones that give the dominant contribution to the mean metric of the
Universe. Moreover, the consistency condition that the expansion rate
obtained with such a mean metric be equal to $H_0$, determines the
value of a corresponding effective density, which is about five times
the observed barionic one, i.e., about equal to the estimated density
of the dark matter.

\section{Form of the force due to the far away galaxies} 
So far for what concerns the mean metric. We now come to the problem
of estimating the effects of the fluctuations on the dynamics of a
test particle. The force per unit mass on a test particle is obtained
in the familiar way through the equation for the geodesics.  Notice
that the Hubble relation (\ref{hubble}) has here an essential
impact. Indeed, the force contains both a term decreasing as $1/r^2$,
which is proportional to the velocity of the source, and a term
decreasing as $1/r$, which is proportional to the acceleration of the
source. Thus, estimating the acceleration too through Hubble's law,
the latter term actually turns out not to depend on distance at all,
and thus it is again the far away matter that is found to give the
dominant contribution.  Compare this with the way in which Mach's
principle was dealt with in \cite{einstein} (see page 102). There,
lacking Hubble's law, the velocities of the sources were
neglected. Thus, only the Newtonian, fast decaying, potential was
considered, and so only the near matter, and not the far one, appeared
to play a role.

So we address our attention to the dominant term of the gravitational
force per unit mass, namely, the one proportional to the acceleration
of the source. Such a term, which we denote by $\vett f$, has, at the
origin, the form
\begin{equation}\label{vettore}
\vett f= \frac {4G H_0^2\, M}{c^2}\ \vett u\ ,\quad \quad \vett u(N)=
\sum_{j=1}^N \frac{\vett q_j}{|\vett q_j|}\
\end{equation}
with the positions ${\vett q}_j$ of the $N$ galaxies are evaluated at
corresponding retarded times. Here, the masses of the galaxies were
all put equal to a common value $M$, and the Lorentz factors
$\gamma_j$ were put equal to 1, for the reasons to be illustrated
later.  So, apart from a multiplicative factor, such a force just
reduces to the sum of the unit vectors pointing to each of the
galaxies at the corresponding retarded time.  Actually, our attention
was addressed to the component of such a force $\vett f$ along a given
direction. Such a component will be simply denoted by $f$, and the
corresponding component of $\vett u$ by $u$.

\section{Estimate of the force. Role of the probabilistic assumptions 
for the distribution of galaxies}
Having determined the quantity $f$ of interest (or equivalently $u$),
we now come to the problem of how to describe the distribution of the
galaxies.  It is immediately seen that $f$ exactly vanishes (at any
point) if the matter is described as a continuous medium with a
spherically symmetric density.  From the probabilistic point of view
considered here, such a result (the vanishing of $f$ for a spherically
symmetric matter density) now reads as the vanishing of the mean value
of $f$ for a spherically symmetric probability density of the position
of a galaxy.

We thus come to an estimate of the variance of the force $f$ (or of
$u$). It will be seen that the result depends on the further
assumptions one introduces concerning the spatial distribution of the
galaxies.  Assume first that the positions $\vett q_j$ of the $N$
galaxies are independent random variables, uniformly distributed with
respect to the Lebesgue measure. Then the sum defining $u$ is found to
grow as $\sqrt N$, just in virtue of the central limit theorem.  For
what concerns the estimate of the force on a test particle, one easily
sees that with the present assumption it is completely negligible,
just because the considered sum behaves as $\sqrt N$ rather than as
$N$ (see later).

So we modify such an assumption and, following all the previously
 mentioned authors, we consider the case in which the position vectors
 of the galaxies present a correlation, i.e., are no more
 independently distributed. Thus, the sum defining $u$ is no more
 constrained to grow as $\sqrt N$, and can have a faster growth.  Just
 for the sake of concreteness, we fix our model by requiring that the
 probability density corresponds to a fractal of dimension $2$.  In
 such a way, however, the analytical computation of the probability
 distribution of the force becomes a quite nontrivial task with
 respect to the Poissonian case considered by Chandrasekhar and von
 Neumann, and also with respect to the fractal, but purely Newtonian,
 case considered by Gabrielli et al. in the papers \cite{roma2} and
 \cite{roma3}.  So we are forced, at least provisionally, to
 investigate the problem by numerical methods.

We proceeded as follows.  In order to estimate the sum defining $u$,
the positions of the $N$ galaxies were extracted (with the method
described in \cite{mandel}) in such a way that the mass distribution
has fractal dimension $2$.  We then studied the dependence of $u$ on
the number $N$ of galaxies, which was made to vary in the range
$1000\le N\le 512,000$, the density being kept constant.  This means
that the positions of the $N$ points were taken to lie inside a cutoff
sphere whose volume was made to increase as $N$. For the values of $N$
investigated, the corresponding radius turns out to be so small with
respect to the present horizon, that the Lorentz factors $\gamma$
could altogether be put equal to $1$ (as was previously assumed), and
more in general the special relativistic character of our model was
actually justified.

\begin{figure}
 \begin{center}
  \includegraphics[width= \textwidth]{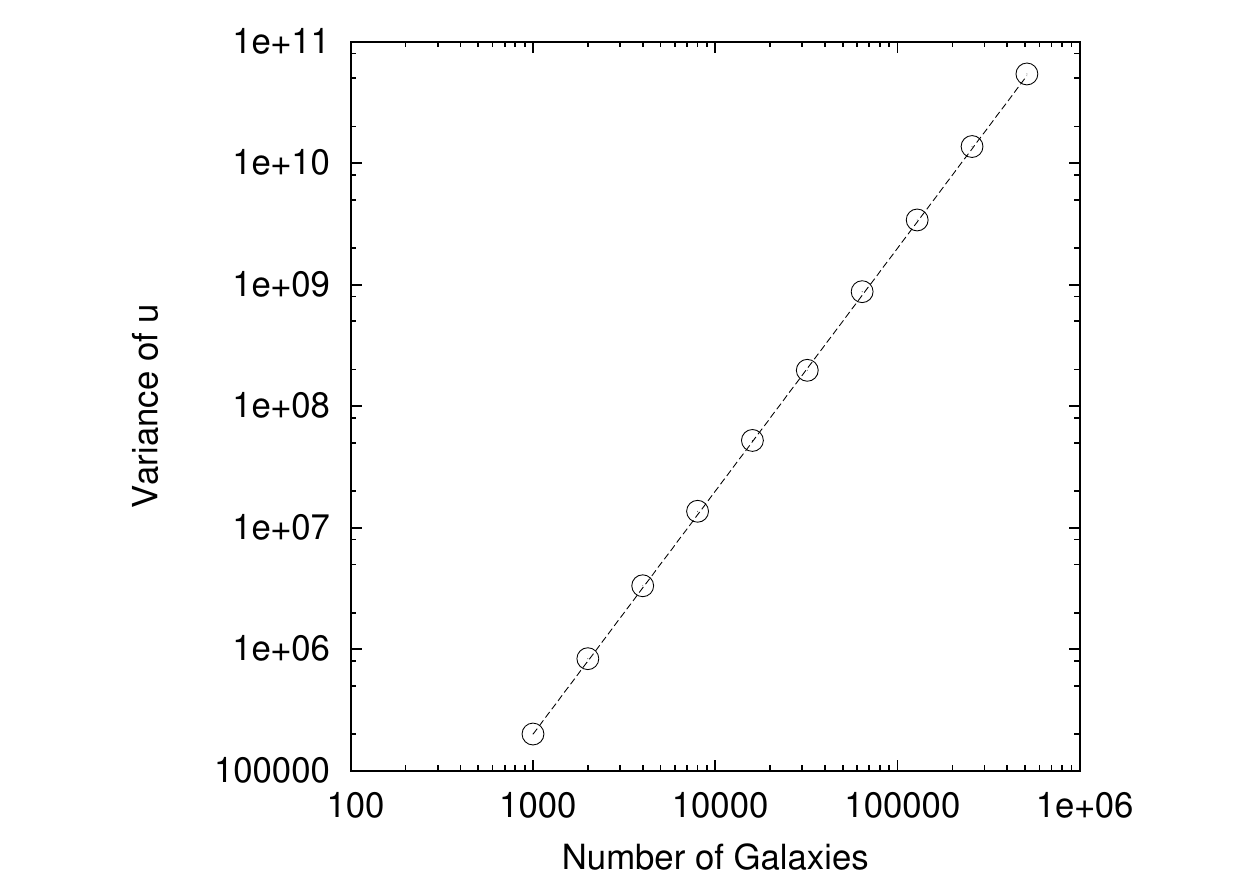}
 \end{center}
 \caption{\label{fig2} The variance $\sigma^2_u$ of $u$ versus the
  number $N$ of galaxies in log--log scale. The dashed line is the
  curve $\sigma_u^2=0.2\ N^2$.}
\end{figure}

The mean of $u$ turns out to practically vanish for all $N$, while its
variance $\sigma^2_u$ is found to grow as $N^2$ (actually, as $0.2\,
N^2$), rather than as $N$, as occurs in the uniform case. This is
shown in Fig.~\ref{fig2}.  The standard deviation $\sigma _f$ is thus
proportional to $N$, being given by
\begin{equation}\label{sigmaf}
\sigma_f\simeq \sqrt{0.2}\ \frac {4GH_0^2}{c^2}\ MN\ =\sqrt{0.2}\
\frac {4G}{R_0^2}\ MN \ .
\end{equation}  

We now take such a result, which was obtained for extremely small
values of $N$, and extrapolate it up to the present horizon
$R_0=c/H_0$, i,e., we insert in formula (\ref{sigmaf}) the actual
value of $N$, so that the quantity $MN$ can be identified with the
total visible mass of the Universe.  The latter can be written as $MN=
(4/3)\, \pi\ \rho_{\eff}\ R_0^3$, in terms of the effective density
$\rho_{\eff}$ previously discussed.  This gives $\sigma_f\simeq 0.2\,
cH_0$.

On the other hand, if a random variable $f$ has zero mean and a finite
variance $\sigma^2_f$, with great probability its modulus will take on
values very near to its standard deviation $\sigma_f$.  In such a
sense we may say to have found
\begin{equation}\label{forza}
|f|\simeq 0.2 \, cH_0\ ,
\end{equation}
which constitutes the second result of the present work.  Namely, in
our oversimplified model the force per unit mass, i.e., the
acceleration, exerted by the far matter on a test particle, is found
to have a value of the order of $cH_0$, which is the one that is met
in most cases in which the presence of a dark matter is
advocated. Notice that the assumption of a uniform, rather than
correlated, distribution of matter would lead instead to $|f|\simeq
cH_0 /{\sqrt N}$, i.e., essentially to $f\simeq 0$. Namely, as
previously mentioned, without the correlation hypothesis for the
positions of the galaxies, the usual procedure of neglecting at all
the gravitational contribution of the far away matter, would be
justified. We expect that the coefficient $\sqrt{0.2}$ in
(\ref{sigmaf}) may depend on the degree of correlation chosen for the
positions of the galaxies. This point will be investigated elsewhere.

Notice that this force, acting on each test particle, also acts on
each galaxy itself, thus producing an acceleration which should be
added to the one given by Hubble's law. On the other hand, such an
acceleration was neglected in our model, because the peculiar
velocities were assumed to vanish, so that we have here a consistency
problem. In this connection we notice that this acceleration is small
with respect the Hubble acceleration $H_0c$ of the far away galaxies
(which are the relevant ones), so that our procedure seems to be
consistent. A more accurate discussion of this point is left for
future work.

\section{Possible application to the virial of the forces for 
a cluster of galaxies}
We now address the problem whether the previous result may be applied
to estimating the virial of the external forces for a cluster of
galaxies.  We have in mind the work of Zwicky \cite{zwi} for the Coma
cluster, in which the contribution of the internal matter was found to
be negligible, that of the external galaxies was not even mentioned
(perhaps, in the spirit of the continuum approximation), and the
presence of a dark matter was proposed.  Let us recall that, according
to the virial theorem, for a confined system constituted by $n$
particles (think of a cluster of galaxies) one has $\overline
{\sigma^2_v} = - \overline \vir\, /n$.  Here, $\sigma^2_v=(1/n)\,
\sum_i v_i^2$ is the variance of the velocity distribution of the
particles (the galaxies of the cluster), whereas $\vir=\sum_i \vett
f_i\cdot \vett x_i$ is called the virial of the forces (per unit
mass), $\vett x_i$ denoting the position vector of the $i$--th
internal particle with respect to the center of mass of the cluster,
while overline denotes time--average. It was shown by Zwicky that the
contribution of the internal forces is negligible, so that in
estimating the virial we can just consider the force due to the
external galaxies.

It is well known that the virial of the external forces (per unit
mass) equals the virial of the tidal force (per unit mass) $f-f^*$
where $f^*$ is the force (per unit mass) at the center of mass,
because the contribution of $f^*$ vanishes. The key point now is that
the contribution of the tidal forces depends on the smoothness
properties of the field of force $f$. Indeed it turns out that, if the
field is smooth, so that the tidal force can be estimated through a
Taylor expansion, then one finds $\overline \vir\, /n\simeq H_0^2L^2$
where $L$ is the linear dimension of the cluster. For the Coma cluster
this contribution turns out to be negligible.

If one instead assumes that the forces at different points be
uncorrelated, then it turns out that the contribution may be of the
correct order of magnitude.

Indeed this assumption has two deep consequences. The first one is
that it makes conceivable that locally, in some regions, the random
field of force may form patterns of a central--like type, which are
attractive towards a center, with a nonvanishing force at the
center. By the way, this is equivalent to the fact that locally, in
such special regions, the external far away matter produces a
pressure.  The second consequence is that in such a case the variance
of the tidal force $f-f^*$ just equals $\sqrt {2}$ the variance of the
force $f$, the estimate of which was given in formula (\ref{forza}).

So, having assumed that the tidal force be of central--like type, the
terms of the sum $\sum_{i=1}^n \overline {(\vett f_i-\vett f^*)\cdot
\vett x_i}$ can be estimated as $\overline {(\vett f_i-\vett f^*)\cdot
\vett x_i}\simeq - \sqrt{2}|f|\ \overline {|\vett x_i|} $, with $f$
given by (\ref{forza}), and with $\overline {|\vett x_i|}\simeq L/4$,
where $L$ is the diameter of the cluster.  So, for the velocity
variance one gets
\begin{equation}
\label{varianza}
\overline{\sigma_v^2}\simeq \sqrt {2}\, 0.05\ { cH_0 L}\simeq 0.07\ {
  cH_0 L} \ .
\end{equation}
In the case of Coma one thus finds a value $\simeq 8 \cdot 10^5
\mathrm{km}^2/\mathrm{sec}^2$, which is very near to the value $5
\cdot 10^5 \, \mathrm{km}^2/\mathrm{sec}^2$ reported by Zwicky.

The prediction that the velocity variance depends linearly on $L$,
 according to (\ref{varianza}), may be of interest, and apparently is
 in agreement with the observations (see \cite{kazanas}, Fig. 2, page
 539, and \cite{combes}).  Notice that, with the parameters entering
 the problem, the square of a velocity can be formed only as $c^2$, or
 as $cH_0L$ or as $(H_0L)^2$. But the first term is by far too large,
 the last term (as previously pointed out) by far too small, while the
 term linear in $L$ is indeed about of the correct order of magnitude.
 Thus, the previous considerations appear to indicate that the
 decorrelation assumption for the forces at different points is
 necessary in order that the observed velocity dispersion in a cluster
 may be ascribed to the gravitational action of the far away galaxies.

We now briefly address the question of understanding which mechanisms
might be reponsible for such a decorrelation. We have in mind two
mechanisms. The first one is suggested by the analysis made in the
paper \cite{roma} of Joyce et al., in which the Newtonian contribution
to the tidal force is estimated, albeit in a different context. Indeed
in such a paper it is shown (see page 418) that the Newtonian
contribution to the tidal force is finite, whereas the purely
Newtonian non tidal contribution would be divergent, at least for
certain values of the fractal dimension. On the other hand, the latter
quantity is just of the same order of magnitude of the tidal force
corresponding to our far fields, and so such a result suggests that
the tidal force due to the far fields may be divergent. This in turns
may be considered as an indication of decorrelation.  The second
mechanism has instead a cosmological character, and is related on the
one hand to the fact that the cosmological horizons relative to
different galaxies do not coincide, and on the other hand to the fact
that the main contribution to the force comes from the matter near the
horizon. Remarking in addition that the distributions of matter about
two different horizons should be considered as independent ones (the
horizons being non causally connected), one is led to conceive that
also the corresponding forces might be independent.

A consistent discussion of this point would require the consideration
of a more realistic model, in which the time dependence of Hubble's
constant be taken into account. So we leave a discussion of this point
for future work.

\section{A comment on the  perturbation approach}
In the present paper we have chosen a perturbation approach in which
the zero--th order solution is the flat metric, and consistently the
first--order correction was found to be small with respect to the
unperturbed one.

One might imagine that a better approximation be obtained if the mean
 metric, i.e., the Friedmann--Robertson--Walker one, is taken directly
 as zero--th order solution. One easily sees, however, that such an
 approximation scheme meets with two difficulties. The first one is
 that in such a case the source entering the equation for the first
 order solution is of the same order as the zero--th order
 source. Indeed, the source is proportional to the quantity
$$ T^{\mu\nu}-<T^{\mu\nu}>
$$ which is not small, as its modulus is almost everywhere equal to
that of $<T^{\mu\nu}>$. A more serious difficulty is the fact that the
first--order perturbation has to satisfy essentially d'Alembert's
equation with the source just mentioned, while such an equation cannot
be solved by elementary methods, and it is not even known whether it
admits bounded solutions at all. So, in paper \cite{dp}, Davis and
Peebles, who use such a perturbation procedure for the analogous
nonrelativistic case, have to introduce a suitable resummation
procedure. Now it is not clear whether a similar resummation procedure
can be introduced also in our relativistic case, and furthermore in
our case a discussion of the boundedness of the solution would be
required because, at variance with Davis and Peebles, we are not
restricting ourselves to the case of short distances.

Now, neither of the mentioned difficulties comes in with our
perturbation procedure.  In addition, it seems to us that the
procedure of David and Peebles eventually is equivalent to the
nonrelativistic version of our procedure.  Indeed, their formula (12)
is equivalent to our formula (\ref{aaaa}), taken in the
nonrelativistic approximation, while their formula (14) just gives the
contribution to the force due to the near galaxies. This contribution
occurs also in our case, and does not appear explicitly in our formula
(\ref{vettore}), only because in the latter we just retained the
dominant contribution due to the far away galaxies.

It is worth mentioning that a perturbation about the FRW metric is
performed also be Joyce et al. in the paper \cite{roma}. But in their
case they take into account the fact that the zero--th order solution
is due to the radiation energy density, so that the energy--momentum
tensor due to matter is not a perturbation of the vacuum. Thus they do
not meet with the previously mentioned problems.

\section{Conclusions}
In conclusion, we have studied the retarded gravitational action of
the far away galaxies. Such an action vanishes if the matter in the
Universe is described in terms of a continuous spherically symmetric
continuum. We have pointed out that such an action is instead quite
relevant if the discrete character of matter, as constituted of
galaxies with correlated positions, is taken into account. Some
gravitational effects were estimated, and were found to have the same
order of magnitude as the corresponding local ones of dark matter.

It is sometimes stated \cite{nature} that the fractal picture of the
Universe may be incompatible with the framework of the standard
cosmological theories, and in the paper \cite{roma} by Joyce et al. a
solution was proposed, based on the idea that the contribution of
matter to Einstein's equations should be considered as a perturbation
to the contribution of radiation. Perhaps the present approach, in
which a perturbation to the vacuum is performed, and the FRW metric is
obtained in the mean (even if radiation is altogether neglected), may
be considered as providing an alternative complementary solution to
the problem.

\vskip .5truecm
\noindent
\textbf{Acknowledgments}. We thank George Contopoulos, Christos
Efthy\-mio\-pou\-los,
Francesco Sylos Labini and Rudolf Thun for useful discussions. We
thank an anonymous referee for pointing out to us that what matters in
the virial theorem for a non--isolated system is the tidal force, and
for drawing our attention to the paper \cite{roma} that had escaped
us, where the Newtonian contribution to the tidal force is estimated
for a fractal distribution of matter. This paper is dedicated to the
memory of Nikos Voglis.

\end{document}